%% file: nEDMLetter.tex
\documentclass[aps,prl,superscriptaddress,showkeys]{revtex4}

\input{packages}

\input{CommandsAndShortcuts.tex}
\newcommand{\del}[1]{}
\newcommand{\R}{\mathcal{R}}
\newcommand{\n}{{\rm n}}
\newcommand{\Hg}{{\rm Hg}}

\begin{document}

\title{Measurement of the permanent electric dipole moment of the neutron}

\author{C.~Abel}
\affiliation{Department of Physics and Astronomy, University of Sussex, Falmer, Brighton BN1 9QH, UK}

\author{S.~Afach}
\affiliation{Paul Scherrer Institut, CH-5232 Villigen PSI, Switzerland}
\affiliation{ETH Z\"{u}rich, Institute for Particle Physics and Astrophysics, CH-8093 Z\"{u}rich, Switzerland}
%\affiliation{Hans Berger Department of Neurology, Jena University Hospital, D-07747 Jena, Germany}
%ETH Zürich, Institute for Particle Physics and Astrophysics, CH-8093 Zürich, Switzerland
%
\author{N.\,J.~Ayres}
\affiliation{Department of Physics and Astronomy, University of Sussex, Falmer, Brighton BN1 9QH, UK}
\affiliation{ETH Z\"{u}rich, Institute for Particle Physics and Astrophysics, CH-8093 Z\"{u}rich, Switzerland}
%
%\author{F.~Atchison}
%\affiliation{Paul Scherrer Institut, CH-5232 Villigen PSI, Switzerland}
%
\author{C.\,A.~Baker}
\affiliation{STFC Rutherford Appleton Laboratory, Harwell, Didcot, Oxon OX11 0QX, United Kingdom}
\author{G.~Ban}
\affiliation{LPC Caen, ENSICAEN, Universit\'{e} de Caen, CNRS/IN2P3, Caen, France}
\author{G.~Bison}
\affiliation{Paul Scherrer Institut, CH-5232 Villigen PSI, Switzerland}
\author{K.~Bodek}
\affiliation{Marian Smoluchowski Institute of Physics, Jagiellonian University, 30-059 Cracow, Poland}
\author{V.~Bondar}
\affiliation{Paul Scherrer Institut, CH-5232 Villigen PSI,
Switzerland}
\affiliation{ETH Z\"{u}rich, Institute for Particle Physics and Astrophysics, CH-8093 Z\"{u}rich, Switzerland}
\affiliation{Instituut voor Kern- en Stralingsfysica, University of Leuven, B-3001 Leuven, Belgium}

\author{M.~Burghoff}
\affiliation{Physikalisch Technische Bundesanstalt, Berlin, Germany}
\author{E.~Chanel}
\affiliation{Laboratory for High Energy Physics and Albert Einstein Center for Fundamental Physics, University of Bern, CH-3012 Bern, Switzerland}
\author{Z.~Chowdhuri}

\affiliation{Paul Scherrer Institut, CH-5232 Villigen PSI, Switzerland}
\author{P.-J.~Chiu}
\affiliation{Paul Scherrer Institut, CH-5232 Villigen PSI, Switzerland}
\affiliation{ETH Z\"{u}rich, Institute for Particle Physics and Astrophysics, CH-8093 Z\"{u}rich, Switzerland}
\author{B.~Clement}
\affiliation{Univ. Grenoble Alpes, CNRS, Grenoble INP, LPSC-IN2P3, Grenoble, France}
\author{C.\,B.~Crawford} 
\affiliation{University of Kentucky, Lexington, USA}
\author{M.~Daum}
\affiliation{Paul Scherrer Institut, CH-5232 Villigen PSI, Switzerland}
\author{S.~Emmenegger}
\affiliation{ETH Z\"{u}rich, Institute for Particle Physics and Astrophysics, CH-8093 Z\"{u}rich, Switzerland}
\author{L.~Ferraris-Bouchez}
\affiliation{Univ. Grenoble Alpes, CNRS, Grenoble INP, LPSC-IN2P3, Grenoble, France}
\author{M.~Fertl}
\affiliation{Paul Scherrer Institut, CH-5232 Villigen PSI, Switzerland}
\affiliation{ETH Z\"{u}rich, Institute for Particle Physics and Astrophysics, CH-8093 Z\"{u}rich, Switzerland}
\affiliation{Institut f\"{u}r Physik, Johannes-Gutenberg-Universit\"{a}t, D-55128 Mainz, Germany}
\author{P.~Flaux}
\affiliation{LPC Caen, ENSICAEN, Universit\'{e} de Caen, CNRS/IN2P3, Caen, France}
\author{B.~Franke}
\altaffiliation[Present address: ]{TRIUMF, Vancouver, Canada.}
\affiliation{Paul Scherrer Institut, CH-5232 Villigen PSI, Switzerland}
\affiliation{ETH Z\"{u}rich, Institute for Particle Physics and Astrophysics, CH-8093 Z\"{u}rich, Switzerland}
\author{A.~Fratangelo}
\affiliation{Laboratory for High Energy Physics and Albert Einstein Center for Fundamental Physics, University of Bern, CH-3012 Bern, Switzerland}
\author{P.~Geltenbort}
\affiliation{Institut Laue-Langevin, CS 20156 F-38042 Grenoble Cedex 9, France}
\author{K.~Green}
\affiliation{STFC Rutherford Appleton Laboratory, Harwell, Didcot, Oxon OX11 0QX, United Kingdom}
\author{W.\,C.~Griffith}
\affiliation{Department of Physics and Astronomy, University of Sussex, Falmer, Brighton BN1 9QH, UK}
\author{M.~van\,der\,Grinten}
\affiliation{STFC Rutherford Appleton Laboratory, Harwell, Didcot, Oxon OX11 0QX, United Kingdom}
\author{Z.\,D.~Gruji\'c}
\affiliation{Physics Department, University of Fribourg, CH-1700 Fribourg, Switzerland}
\affiliation{Institute of Physics Belgrade, University of Belgrade, 11080 Belgrade, Serbia}
\author{P.\,G.~Harris}
\affiliation{Department of Physics and Astronomy, University of Sussex, Falmer, Brighton BN1 9QH, UK}
\author{L.~Hayen}
\altaffiliation[Present address: ]{Department of Physics, North Carolina State University, Raleigh, NC 27695, USA}
\affiliation{Instituut voor Kern- en Stralingsfysica, University of Leuven, B-3001 Leuven, Belgium}
\author{W.~Heil}
\affiliation{Institut f\"{u}r Physik, Johannes-Gutenberg-Universit\"{a}t, D-55128 Mainz, Germany}
\author{R.~Henneck}
\affiliation{Paul Scherrer Institut, CH-5232 Villigen PSI, Switzerland}
\author{V.~H\'{e}laine}
\affiliation{Paul Scherrer Institut, CH-5232 Villigen PSI, Switzerland}
\affiliation{LPC Caen, ENSICAEN, Universit\'{e} de Caen, CNRS/IN2P3, Caen, France}
\author{N.~Hild}
\affiliation{Paul Scherrer Institut, CH-5232 Villigen PSI, Switzerland}
\affiliation{ETH Z\"{u}rich, Institute for Particle Physics and Astrophysics, CH-8093 Z\"{u}rich, Switzerland}
\author{Z.~Hodge}
\affiliation{Laboratory for High Energy Physics and Albert Einstein Center for Fundamental Physics, University of Bern, CH-3012 Bern, Switzerland}
\author{M.~Horras}
\affiliation{Paul Scherrer Institut, CH-5232 Villigen PSI, Switzerland}
\affiliation{ETH Z\"{u}rich, Institute for Particle Physics and Astrophysics, CH-8093 Z\"{u}rich, Switzerland}
\author{P.~Iaydjiev}
\altaffiliation[On leave from ]{Institute of Nuclear Research and Nuclear
Energy, Sofia, Bulgaria.}
\affiliation{STFC Rutherford Appleton Laboratory, Harwell, Didcot, Oxon OX11 0QX, United Kingdom}
\author{S.\,N.~Ivanov}
\altaffiliation[On leave from ]{Petersburg Nuclear Physics Institute, Russia.}
\affiliation{STFC Rutherford Appleton Laboratory, Harwell, Didcot, Oxon OX11 0QX, United Kingdom}
\author{M.~Kasprzak}
\affiliation{Paul Scherrer Institut, CH-5232 Villigen PSI, Switzerland}
\affiliation{Instituut voor Kern- en Stralingsfysica, University of Leuven, B-3001 Leuven, Belgium}
\author{Y.~Kermaidic}
\altaffiliation[Present address:  ]{Max-Planck-Institut fur Kernphysik, Heidelberg, Germany.}
\affiliation{Univ. Grenoble Alpes, CNRS, Grenoble INP, LPSC-IN2P3, Grenoble, France}
\author{K.~Kirch}
\affiliation{Paul Scherrer Institut, CH-5232 Villigen PSI,
Switzerland}
\affiliation{ETH Z\"{u}rich, Institute for Particle Physics and Astrophysics, CH-8093 Z\"{u}rich, Switzerland}
\author{A.~Knecht}
\affiliation{Paul Scherrer Institut, CH-5232 Villigen PSI,
Switzerland}
\affiliation{ETH Z\"{u}rich, Institute for Particle Physics and Astrophysics, CH-8093 Z\"{u}rich, Switzerland}
\author{P.~Knowles}
\affiliation{Physics Department, University of Fribourg, CH-1700 Fribourg, Switzerland}
\author{H.-C.~Koch}
\affiliation{Paul Scherrer Institut, CH-5232 Villigen PSI, Switzerland}
\affiliation{Physics Department, University of Fribourg, CH-1700 Fribourg, Switzerland}
\affiliation{Institut f\"{u}r Physik, Johannes-Gutenberg-Universit\"{a}t, D-55128 Mainz, Germany}
\author{P.A.~Koss}
\altaffiliation[Present address: ]{Fraunhofer-Institut f\"{u}r Physikalische Messtechnik IPM, 79110 Freiburg i. Breisgau, Germany}
\affiliation{Instituut voor Kern- en Stralingsfysica, University of Leuven, B-3001 Leuven, Belgium}
\author{S.~Komposch}
\affiliation{Paul Scherrer Institut, CH-5232 Villigen PSI,
Switzerland}
\affiliation{ETH Z\"{u}rich, Institute for Particle Physics and Astrophysics, CH-8093 Z\"{u}rich, Switzerland}
\author{A.~Kozela}
\affiliation{Henryk Niedwodniczanski Institute for Nuclear Physics, Cracow, Poland}
\author{A.~Kraft}
\affiliation{Paul Scherrer Institut, CH-5232 Villigen PSI, Switzerland}
\affiliation{Institut f\"{u}r Physik, Johannes-Gutenberg-Universit\"{a}t, D-55128 Mainz, Germany}
\author{J.~Krempel}
\affiliation{ETH Z\"{u}rich, Institute for Particle Physics and Astrophysics, CH-8093 Z\"{u}rich, Switzerland}
\author{M.~Ku\'zniak}
\altaffiliation[Present address: ]{AstroCeNT, Nicolaus Copernicus Astronomical Center, Polish Academy of Sciences, Rektorska 4, Warsaw, Poland}
\affiliation{Paul Scherrer Institut, CH-5232 Villigen PSI, Switzerland}
\affiliation{Marian Smoluchowski Institute of Physics, Jagiellonian University, 30-059 Cracow, Poland}
\author{B.~Lauss}
\affiliation{Paul Scherrer Institut, CH-5232 Villigen PSI,
Switzerland}
\author{T.~Lefort}
\author{Y.~Lemi\`{e}re}
\affiliation{LPC Caen, ENSICAEN, Universit\'{e} de Caen, CNRS/IN2P3, Caen, France}
\author{A.~Leredde}
\affiliation{Univ. Grenoble Alpes, CNRS, Grenoble INP, LPSC-IN2P3, Grenoble, France}
\author{P.~Mohanmurthy}
\affiliation{Paul Scherrer Institut, CH-5232 Villigen PSI,
Switzerland}
\affiliation{ETH Z\"{u}rich, Institute for Particle Physics and Astrophysics, CH-8093 Z\"{u}rich, Switzerland}
\author{A.~Mtchedlishvili}
\affiliation{Paul Scherrer Institut, CH-5232 Villigen PSI,
Switzerland}
\author{M.~Musgrave}
\altaffiliation[Present address:  ]{MIT, Boston, USA.}
\affiliation{Department of Physics and Astronomy, University of Sussex, Falmer, Brighton BN1 9QH, UK}
\author{O.~Naviliat-Cuncic} 
\affiliation{LPC Caen, ENSICAEN, Universit\'{e} de Caen, CNRS/IN2P3, Caen, France}
\author{D.~Pais}
\affiliation{Paul Scherrer Institut, CH-5232 Villigen PSI,
Switzerland}
\affiliation{ETH Z\"{u}rich, Institute for Particle Physics and Astrophysics, CH-8093 Z\"{u}rich, Switzerland}
\author{F.M.~Piegsa}
\affiliation{Laboratory for High Energy Physics and Albert Einstein Center for Fundamental Physics, University of Bern, CH-3012 Bern, Switzerland}
\author{E.~Pierre}
\altaffiliation[Present address: ]{LPTMC, Sorbonne Universit\'{e}, Paris, France}
\affiliation{Paul Scherrer Institut, CH-5232 Villigen PSI,
Switzerland}
\affiliation{LPC Caen, ENSICAEN, Universit\'{e} de Caen, CNRS/IN2P3, Caen, France}
\author{G.~Pignol}
\email[Corresponding author: ]{pignol@lpsc.in2p3.fr}
\affiliation{Univ. Grenoble Alpes, CNRS, Grenoble INP, LPSC-IN2P3, Grenoble, France}
\author{C.~Plonka-Spehr}
\affiliation{Institut f\"{u}r Kernchemie, Johannes-Gutenberg-Universit\"{a}t, Mainz, Germany}
\author{P.\,N.~Prashanth}
\affiliation{Instituut voor Kern- en Stralingsfysica, University of Leuven, B-3001 Leuven, Belgium}
\author{G.~Qu\'{e}m\'{e}ner}
\affiliation{LPC Caen, ENSICAEN, Universit\'{e} de Caen, CNRS/IN2P3, Caen, France}
\author{M.~Rawlik}
\altaffiliation[Present address:  ]{Paul Scherrer Institut, CH-5232 Villigen PSI, Switzerland}
\affiliation{ETH Z\"{u}rich, Institute for Particle Physics and Astrophysics, CH-8093 Z\"{u}rich, Switzerland}
\author{D.~Rebreyend}
\affiliation{Univ. Grenoble Alpes, CNRS, Grenoble INP, LPSC-IN2P3, Grenoble, France}
\author{I.~Rien\"{a}cker}
\affiliation{Paul Scherrer Institut, CH-5232 Villigen PSI,
Switzerland}
\affiliation{ETH Z\"{u}rich, Institute for Particle Physics and Astrophysics, CH-8093 Z\"{u}rich, Switzerland}
\author{D.~Ries}
\affiliation{Paul Scherrer Institut, CH-5232 Villigen PSI, Switzerland}
\affiliation{ETH Z\"{u}rich, Institute for Particle Physics and Astrophysics, CH-8093 Z\"{u}rich, Switzerland}
\affiliation{Institut f\"{u}r Kernchemie, Johannes-Gutenberg-Universit\"{a}t, Mainz, Germany}
\author{S.~Roccia}
\email[Corresponding author: ]{roccia@ill.fr}
%\altaffiliation[On leave from ]{CSNSM, Universit\'{e} Paris Sud, CNRS/IN2P3, Orsay, France}
\affiliation{Institut Laue-Langevin, CS 20156 F-38042 Grenoble Cedex 9, France}
\affiliation{CSNSM, Universit\'{e} Paris Sud, CNRS/IN2P3, Orsay, France}
\author{G.~Rogel}
\altaffiliation[Present address: ]{CEA Saclay, Saclay, France}
\affiliation{LPC Caen, ENSICAEN, Universit\'{e} de Caen, CNRS/IN2P3, Caen, France}
%
%\author{K.~Ro\ss}
%\affiliation{Institut f\"{u}r Kernchemie, Johannes-Gutenberg-Universit\"{a}t, Mainz, Germany}
%
\author{D.~Rozpedzik}
\affiliation{Marian Smoluchowski Institute of Physics, Jagiellonian University, 30-059 Cracow, Poland}
\author{A.~Schnabel}
\affiliation{Physikalisch Technische Bundesanstalt, Berlin, Germany}
\author{P.~Schmidt-Wellenburg}
\email[Corresponding author: ]{philipp.schmidt-wellenburg@psi.ch} \affiliation{Paul Scherrer Institut, CH-5232 Villigen PSI, Switzerland}
\author{N.~Severijns}
\affiliation{Instituut voor Kern- en Stralingsfysica, University of Leuven, B-3001 Leuven, Belgium}
\author{D.~Shiers}
\affiliation{Department of Physics and Astronomy, University of Sussex, Falmer, Brighton BN1 9QH, UK}
\author{R.~Tavakoli~Dinani}
\affiliation{Instituut voor Kern- en Stralingsfysica, University of Leuven, B-3001 Leuven, Belgium}
\author{J.\,A.~Thorne}
\affiliation{Department of Physics and Astronomy, University of Sussex, Falmer, Brighton BN1 9QH, UK}
\affiliation{Laboratory for High Energy Physics and Albert Einstein Center for Fundamental Physics, University of Bern, CH-3012 Bern, Switzerland}
\author{R.~Virot}
\affiliation{Univ. Grenoble Alpes, CNRS, Grenoble INP, LPSC-IN2P3, Grenoble, France}
\author{J.~Voigt}
\affiliation{Physikalisch Technische Bundesanstalt, Berlin, Germany}
\author{A.~Weis}
\affiliation{Physics Department, University of Fribourg, CH-1700 Fribourg, Switzerland}
\author{E.~Wursten}
\altaffiliation[Present address: ]{CERN, 1211 Gen\`eve, Switzerland}
\affiliation{Instituut voor Kern- en Stralingsfysica, University of Leuven, B-3001 Leuven, Belgium}
\author{G.~Wyszynski}
\affiliation{ETH Z\"{u}rich, Institute for Particle Physics and Astrophysics, CH-8093 Z\"{u}rich, Switzerland}
\affiliation{Marian Smoluchowski Institute of Physics, Jagiellonian University, 30-059 Cracow, Poland}
\author{J.~Zejma}
\affiliation{Marian Smoluchowski Institute of Physics, Jagiellonian University, 30-059 Cracow, Poland}
\author{J.~Zenner}
\affiliation{Paul Scherrer Institut, CH-5232 Villigen PSI,
Switzerland}
\affiliation{Institut f\"{u}r Kernchemie, Johannes-Gutenberg-Universit\"{a}t, Mainz, Germany}
\author{G.~Zsigmond}
\affiliation{Paul Scherrer Institut, CH-5232 Villigen PSI, Switzerland}
%
%\author{A.~Author}
%\affiliation{An institute, somewhere}

\begin{abstract}
We present the result of an experiment to measure the electric dipole moment (EDM) of the neutron  at the Paul Scherrer Institute using Ramsey's method of separated oscillating magnetic fields with ultracold neutrons (UCN). 
Our measurement stands in the long history of EDM experiments probing physics violating time reversal invariance.  %PH: omitted previous sentence since strictly speaking nEDM violates P and T; only CP via the CPT theorem
%Here we report the results from a search for the EDM of the neutron (nEDM) .
%stored for \SI{180}{s} resulting in a resonance width of \SI{2.7}{\milli\hertz}. 
The salient features of this experiment were the use of a \magHg{} co-magnetometer and an array of optically pumped cesium vapor magnetometers to cancel and correct for magnetic field changes. 
%We searched for a change in the ratio of the mercury to neutron precession frequency as function of a strong electric field of \SI{11}{kV/cm} parallel or anti-parallel to a magnetic field of \SI{1}{\muT}. 
The statistical analysis was performed on blinded datasets by two separate groups while the estimation of systematic effects profited from an unprecedented knowledge of the magnetic field. 
The measured value of the neutron EDM is $\dn = \unit[\pow{(0.0\pm1.1_{\rm stat}\pm0.2_{\rm sys})}{-26}]{\ecm}$.
%\it{Add a comment here on the impact of the result.}
\end{abstract}

\maketitle

A nonzero permanent electric dipole moment $\vec{d}\,=\,2d\vec{s}/\hbar$ for a non-degenerate particle with spin $\vec{s}$ implies the violation of time-reversal symmetry. 
Invoking the {\it CPT} theorem\,\cite{Luders1954, Pauli1955} for quantum field theories, this also indicates the  violation of the combined symmetry of charge conjugation and parity ({\it CP}). 
The standard model of particle physics~(SM) contains two sources of {\it CP} violation: the phase of the CKM matrix resulting in the observed {\it CP}-violation in $K$- and $B$-meson decays, and the $\bar{\theta}_{\rm QCD}$ coefficient of the still-unobserved {\it CP}-violating term of the QCD Lagrangian\,\cite{tHooft1976}. 
Both are too small to account for the observed baryon asymmetry of the Universe\,\cite{Morrissey2012NJP} which requires  {\it CP} violation as one of three essential ingredients\,\cite{Sakharov1967}. 
Furthermore, many theories beyond the SM naturally have large {\it CP}-violating phases\,\cite{Engel2013PPNP} that would result in an observable neutron EDM (nEDM)\@. 
In combination with the limits from searches for the  electron\,\cite{Andreev2018} and \magHg\,\cite{Graner2016PRL} EDM, the limit on the nEDM confirms and complements stringent constraints upon many theoretical models\,\cite{Chupp2015PRC}. In particular, the nEDM alone stringently limits $\bar{\theta}_{\rm QCD}$. This unnaturally small upper limit on $\bar{\theta}_{\rm QCD}$ is known as the strong {\it CP} problem; it gave rise to searches for a Goldstone boson, the axion \cite{Wilczek1978,Peccei1977}, which is also an attractive candidate to solve the dark matter mystery \cite{Swart2017}.

\begin{figure}
	\centering
	\includegraphics[width=0.6\columnwidth]{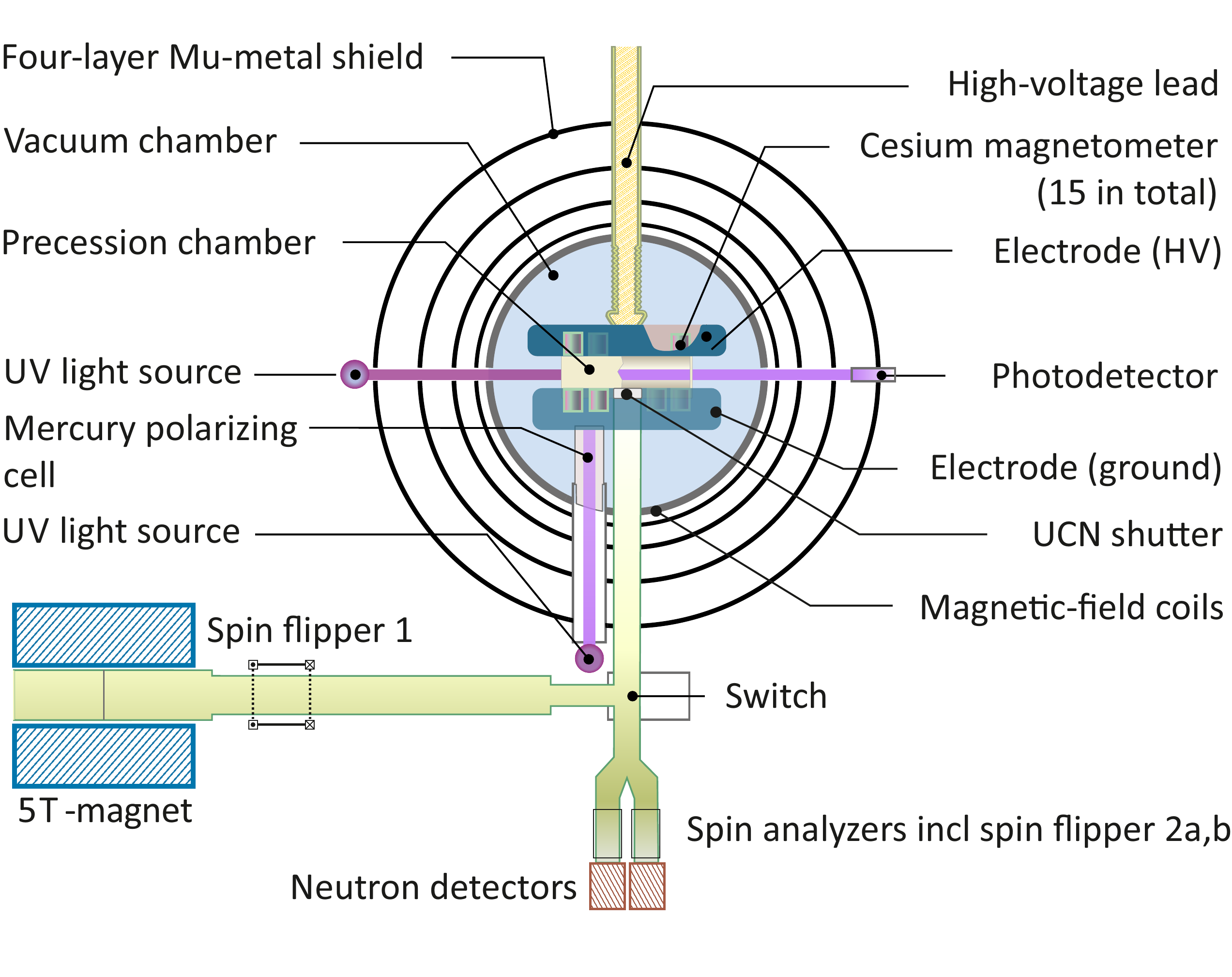}
	\caption{Scheme of the spectrometer used to search for an nEDM. A nonzero signal manifests as shift of the magnetic resonance frequency of polarized UCN in a magnetic field $B_0$ when exposed to an electric field of strength 
	$E$.} 
	\label{fig:apparatus}
\end{figure}

An overview of the spectrometer used for the measurement is shown in Fig.\,\ref{fig:apparatus}, while a detailed technical description of the apparatus (upgraded from that used for the previous best limit\,\cite{Baker2006,Baker2014,Pendlebury2015PRD}) and of data taking may be found in Ref.\,\cite{Abel2018nEDMProc}. 
A total of 54068 individual measurement cycles, during 2015 and 2016, were used to determine the change in the Larmor precession frequency of the neutron,
\begin{equation}
f_{\n} = \frac{1}{\pi \hbar} \left| \mu_{\n} \vec{B_0} + \dn \vec{E} \right|,
\label{eq:EDMequation}
\end{equation}
correlated with the change of polarity of the electric field $|\vec{E}|=\SI{11}{kV/cm}$, where $\mu_{\rm n}$ is the magnetic dipole moment and $\vec{B_0}$ a co-linear magnetic field ($|\vec{B_0}| = 1036 \, \nT$). For this purpose we used Ramsey's method of separate oscillating fields\,\cite{Ramsey1950PR}.

In each cycle UCN from the Paul Scherrer Institute's UCN source\,\cite{Anghel2009,Lauss2014} were polarized by transmission through a \SI{5}{\tesla} superconducting solenoid;  spin flipper 1 (SF1) then allowed the selection of the initial spin state (up or down). 
The switch directed the incoming neutrons to the cylindrical precession chamber 
%(radius $R = \SI{23.5}{cm}$, height $H=\SI{12}{cm}$) 
situated \SI{1.2}{\meter} above the beam line.
The precession chamber (radius $R = \SI{23.5}{cm}$, height $H=\SI{12}{cm}$) was made of diamond like carbon coated\,\cite{Atchison2006PLB,Atchison2006PRC} aluminum electrodes and a deuterated-polystyrene (dPS)\,\cite{Bodek2008} coated insulator ring milled from bulk polystyrene.
After \SI{28}{s} an equilibrium density of up to $\unit[2]{UCN/cm^3}$ inside the precession chamber was attained, and a UCN shutter in the bottom electrode was closed
to confine the UCN for a total of \SI{188}{s}. 
%During \SI{188}{s} the UCN were kept in the precession cell made of DLC coated\,\cite{Atchison2006PLB,Atchison2006PRC} aluminum electrodes and a deuterated polystyrene (dPS)\,\cite{Bodek2008} coated insulator ring milled from bulk polystyrene. 
%Once the UCN were confined a 
A small valve was opened for \SI{2}{s} to release a sample of polarized \magHg{} vapor, that was used as a co-magnetometer (HgM). 
A first low-frequency (LF) pulse of \SI{2}{s} duration and   frequency $|\mu_\Hg B_0|/(\pi\hbar) \approx \SI{7.8}{\hertz}$ tipped the \magHg{} spin by $\pi/2$. 
Ramsey's technique was then applied to the neutrons, with an LF pulse (also of $t_{\rm LF}=\SI{2}{s}$ duration) at a  frequency  $|\mu_\n B_0|/(\pi\hbar) \approx \SI{30.2}{\hertz}$  tipping the UCN spins by $\pi/2$. 
After a period $T= \SI{180}{s}$ of free precession a second neutron LF pulse, in phase with the first,
was applied.
%tipped the UCN spins again. 
%During data taking, four LF frequencies in the steep part of the central Ramsey fringe were alternated.
During data taking, the LF pulses were alternated between four frequencies in the steep regions of the central Ramsey fringe.

Immediately after the second neutron LF pulse the UCN shutter in the bottom electrode was opened. The switch was also moved to the ``empty'' position connecting the precession chamber with the UCN detection system\,\cite{Afach2015EPJA,Ban2016EPJA}, which counted both spin states simultaneously in separate detectors. 
The state of the spin flippers (SF2a/SF2b) above each detector was alternated every fourth cycle, with one of them being off while the other was on, to average over detection, spin flipper, and spin analyzer efficiencies. 
For each cycle $i$, we recorded an asymmetry value between the number of spin up ($N_{{\rm u},i}$) and spin down neutrons ($N_{{\rm d},i}$): $A_i = (N_{{\rm u},i}-N_{{\rm d},i})/(N_{{\rm u},i}+N_{{\rm d},i})$. 
On average, $N_{\rm u}+N_{\rm d} = 11400$ neutrons were counted per cycle. 

In addition, for each cycle we obtained a frequency $f_{{\rm Hg},i}$ from the analysis of the mercury precession signal, as well as 15 frequencies $f_{{\rm Cs},i}$ from cesium magnetometers (CsM) positioned above and below the chamber.
%\footnote{The array was originally made up of 16 cesium magnetometers, unfortunately one broke very early during data taken}. 

There are 22 base configurations of the magnetic field within the dataset. 
Each base configuration was defined by a full degaussing of the four-layer magnetic shield and an ensuing magnetic field optimization using all CsM described in detail in Ref.\,\cite{Elise2019}. 
This procedure was essential to maintain a high visibility, which was measured to be $\overline{\alpha}=0.76$ on average. % note I deleted the "at the end of the precession" because it has to be maintained throughout, and really "measured" is what counts.
A base configuration was kept for a duration of up to a month, during which only the currents of two saddle coils on the vacuum tank, above and below the chamber, were changed to adjust the vertical gradient in a range of approximately $\pm \SI{25}{pT/cm}$\,\cite{Afach2015PRD}. 
Within a base configuration, all cycles with the same applied magnetic gradients were grouped in one sequence. 
The analyzed dataset consists of 99 sequences. 
The voltage applied to the top electrode was changed periodically: eight cycles at zero volts followed by 48 cycles at $\pm 132$~kV, with the pattern then being repeated under reversed polarity.  
During the analysis sequences were split into sub-sequences having polarity patterns of $+--+$ or $-++-$.

\begin{figure}
	\centering
	\includegraphics[width=0.6\columnwidth]{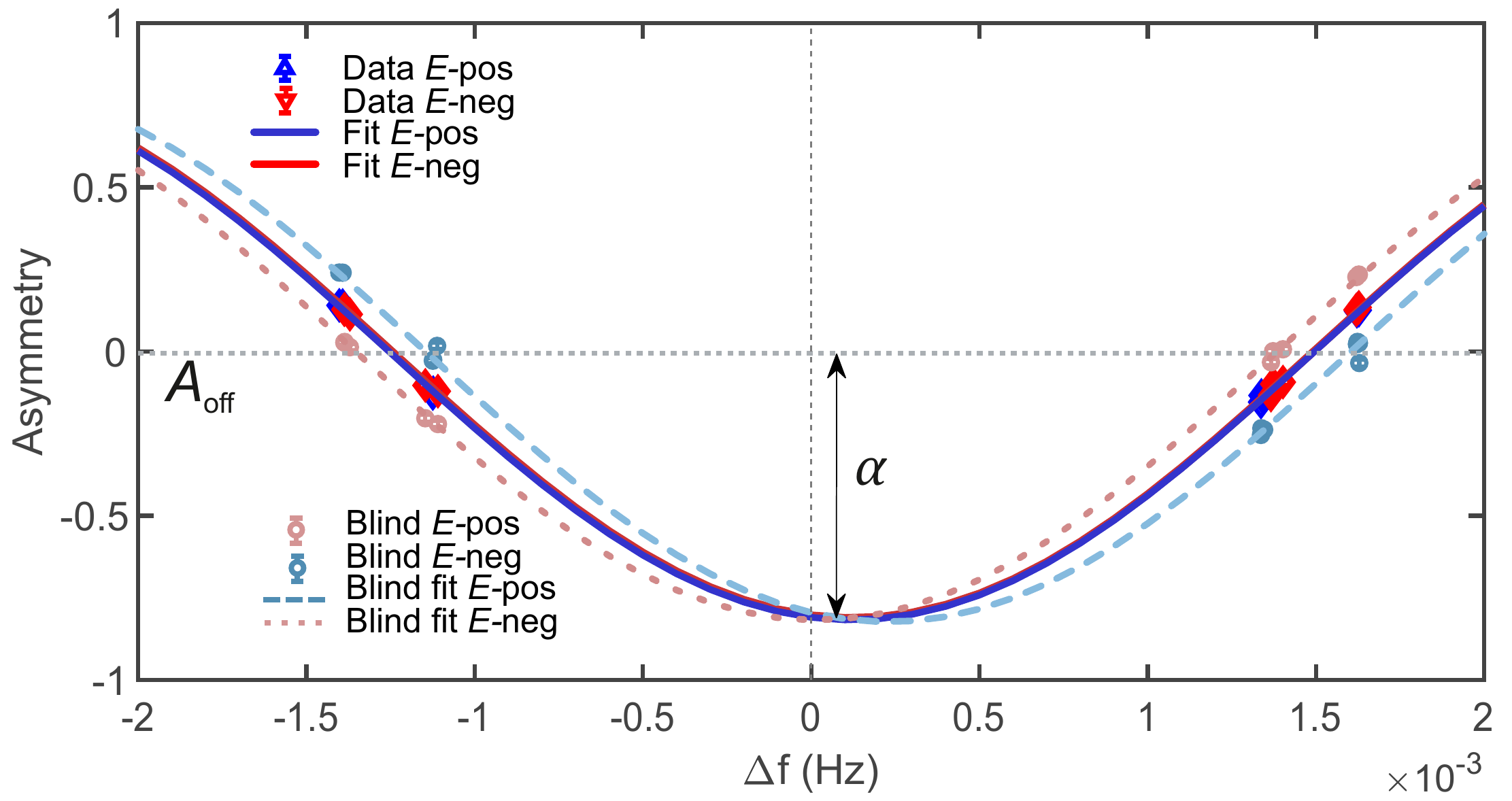}
	\caption{Illustration of the fit to the Ramsey central fringe. Data without electric field are omitted. The data scatters around the four working points. Faded data and lines are for the blinded case (illustration for very large artificial EDM).}
	\label{fig:ramsey_example}
\end{figure}

The analysis searched for shifts in the neutrons' Larmor precession frequency that were proportional to the applied electric field $E_i$. To determine the neutron frequency $f_{\n,i}$ for each cycle from the measured asymmetry $A_i$ 
%by initially fitting  
we fitted the Ramsey resonance
\begin{equation}
			A_i = A_{\rm off} \mp \alpha\cos\left(\frac{\pi\Delta f_i}{\Delta\nu}+\Phi\right)
\label{eq:RamseyResonance}
\end{equation}
to the data of each sub-sequence (see Fig.\,\ref{fig:ramsey_example}), with negative (positive) sign for SF1 turned off (on).
In Eq.\,\eqref{eq:RamseyResonance} $\Delta\nu=(2T+8t_{\rm LF}/\pi)^{-1}=\SI{2.7}{mHz}$ is the resonance linewidth, $\Delta f_i$ is the applied spin-flip frequency $f_{\rm n, LF}$ corrected for magnetic-field changes\,\footnote{One analysis team has chosen to use $\Delta f_i = \langle f_{\rm Hg}\rangle\tfrac{f_{\rm n,LF}}{f_{{\rm Hg},i}}$ and $\Phi=\mathcal{R}\langle f_{\rm Hg}\rangle\tfrac{\Delta\nu}{\pi}$, while the other team used $\Delta f =f_{{\rm Hg},i}-\langle z \rangle g_z$ and the phase $\Phi$.}, and $A_{\rm off}$, $\alpha$, and $\Phi$, are free parameters: offset, fringe visibility, and phase, respectively.
Individual values of $f_{\n,i}$ per cycle were extracted by keeping the fit parameters fixed and rearranging Eq.\,\eqref{eq:RamseyResonance} for $\Delta f_{i}$.

\begin{figure}
	\centering
	\includegraphics[width=0.6\columnwidth]{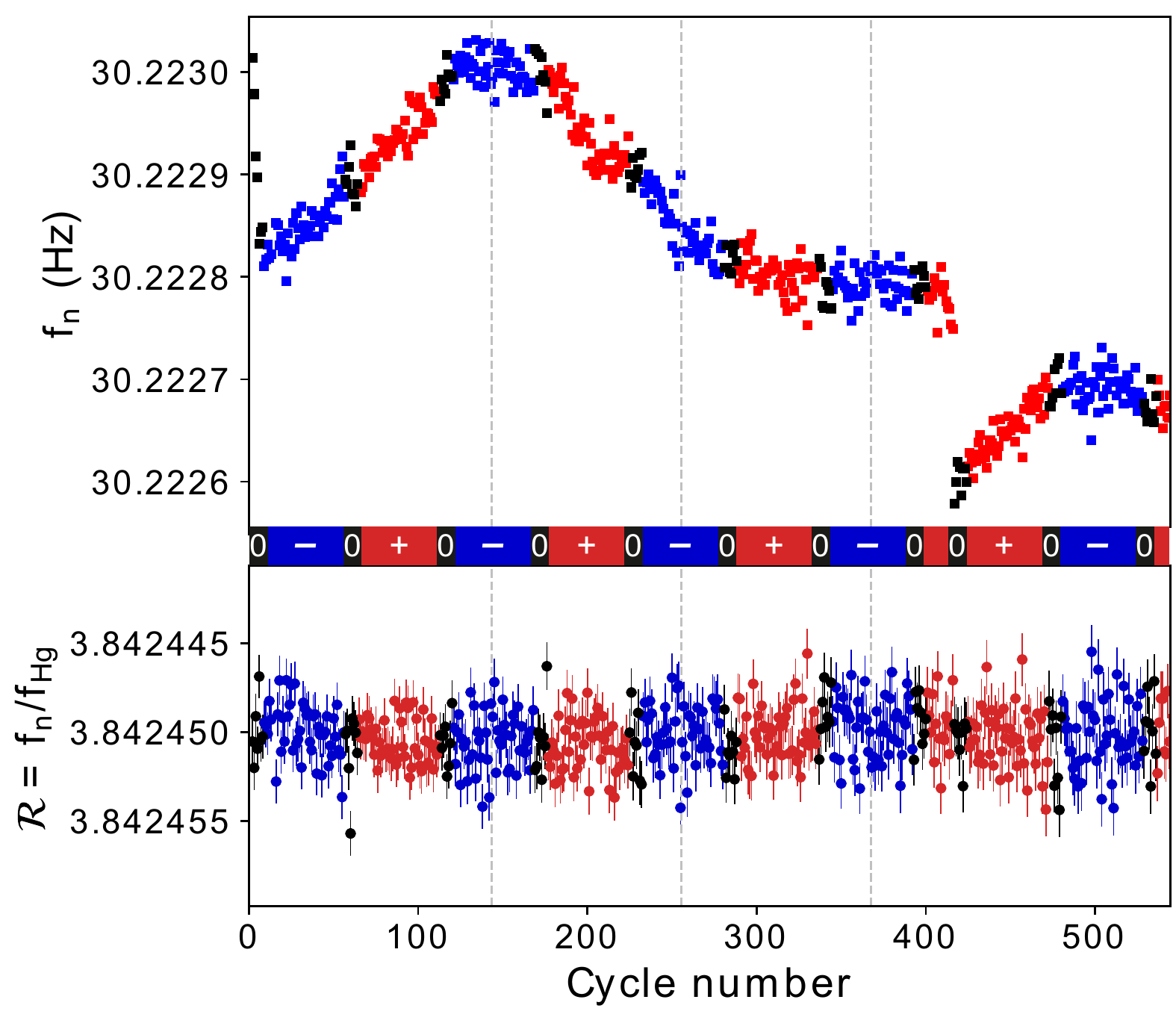}
	\caption{A typical sequence of cycles. Upper plot shows the neutron frequency $f_\n$ as a function of cycle number; lower plot shows the frequency ratio $\R$. The colors correspond to the high-voltage polarity (blue: negative, red: positive, black: zero). The vertical lines separate the sub-sequences. }
	\label{fig:sequence_plot}
\end{figure}

The ratio of frequencies $\R_i = f_{\n,i}/f_{{\rm Hg},i}$ was then used to compensate for residual magnetic-field fluctuations and drifts as shown in Fig.\,\ref{fig:sequence_plot}. 
In what follows, the statistical analysis and the evaluation of systematic effects take into account all known effects affecting the ratio $\R_i$.
These are summarized in the formula
\begin{align}
\label{eq:AllShifts}
	\R = \left|\frac{\gamma_\n}{\gamma_\Hg}\right| &\left(1+\delta_{\rm EDM}\right.  + \delta_{\rm EDM}^{\rm false}+\delta_{\rm quad} \\ \nonumber
							&\left. +\delta_{\rm grav} +\delta_{\rm T} +\delta_{\rm Earth}+\delta_{\rm light}+\delta_{\rm inc}+\delta_{\rm other} \right), 
%			 & +\delta_{\rm earth}+\delta_{\rm light} +\delta_{\rm inc} +\delta_{\rm pulse} +\left. \delta_{\rm ac} \right). \nonumber
\end{align}
where the true EDM term is written 
\begin{equation}
\delta_{\rm EDM} = - \frac{2 E}{\hbar |\gamma_\n| B_0} (\dn + d_{\rm n \leftarrow Hg}) 
\end{equation}
and neglecting the index $i$ for the following.
The \magHg\,EDM, measured to be $d_{\rm Hg} = \unit[\pow{\left(-2.20\pm2.75_{\rm stat}\pm1.48_{\rm sys}\right)}{-30}]{\ecm}$ \,\cite{Graner2016PRL}, induces a bias of the EDM term by $d_{\rm n \leftarrow Hg} = |\gamma_\n/\gamma_\Hg| d_{\rm Hg} =  \unit[\left(-0.1\pm0.1\right)\times 10^{-28}]{\ecm}$, which we quote as a global systematic error. 

Subsequent terms are undesirable effects that influence the neutron or mercury frequencies. We now discuss them individually. 

The gravitational shift $\delta_{\rm grav} = G_{\rm grav} \langle z \rangle / B_0$ induced by the effective vertical magnetic-field  gradient $G_{\rm grav}$ is due to the center of mass offset $\langle z \rangle = -\SI{0.39(3)}{cm}$ of the UCNs in the chamber. 
We deduced $\langle z \rangle$ in an auxiliary analysis from an estimation of the slope $\partial \R/\partial G_{\rm grav}$ by combining the CsM-array readings and offline magnetic-field maps.
The static part of $G_{\rm grav}$ induces a shift of the mean value of $\R$ in a sequence, whereas the fluctuating part induces a drift in $\R$ within each sub-sequence. 
This gradient drift is compensated for at the cycle level using a combination of the HgM and the CsMs below the grounded bottom electrode. The CsMs mounted on the top electrode were not included in order to avoid any possible HV susceptibility in their readings.

In each sub-sequence, we extract the EDM signal $\dn^{\rm meas}$ by fitting the $\R_i$ values, compensated for the gradient drift, as a function of time and electric field, and allowing in addition for a linear time drift. 
% All errors from the initial fit to the Ramsey resonance were propagated by  calculating the explicit covariant matrix for the least square fit. 
%This assumes that all terms in equation \eqref{eq:AllShifts}, except $\delta_{\rm EDM}$ and $\delta_{\rm grav}$, are drifting linearly in time during a sub-sequence and are independent of the electric field. 
This assumes perfect compensation of $\delta_{\rm grav}$, and that $\delta_{\rm EDM}$ is the only $E$-field dependent term in Eq.\,\eqref{eq:AllShifts}. 
Deviations from this hypothesis are treated as systematic effects. 

\begin{figure}
\centering
\includegraphics[width=.6\columnwidth]{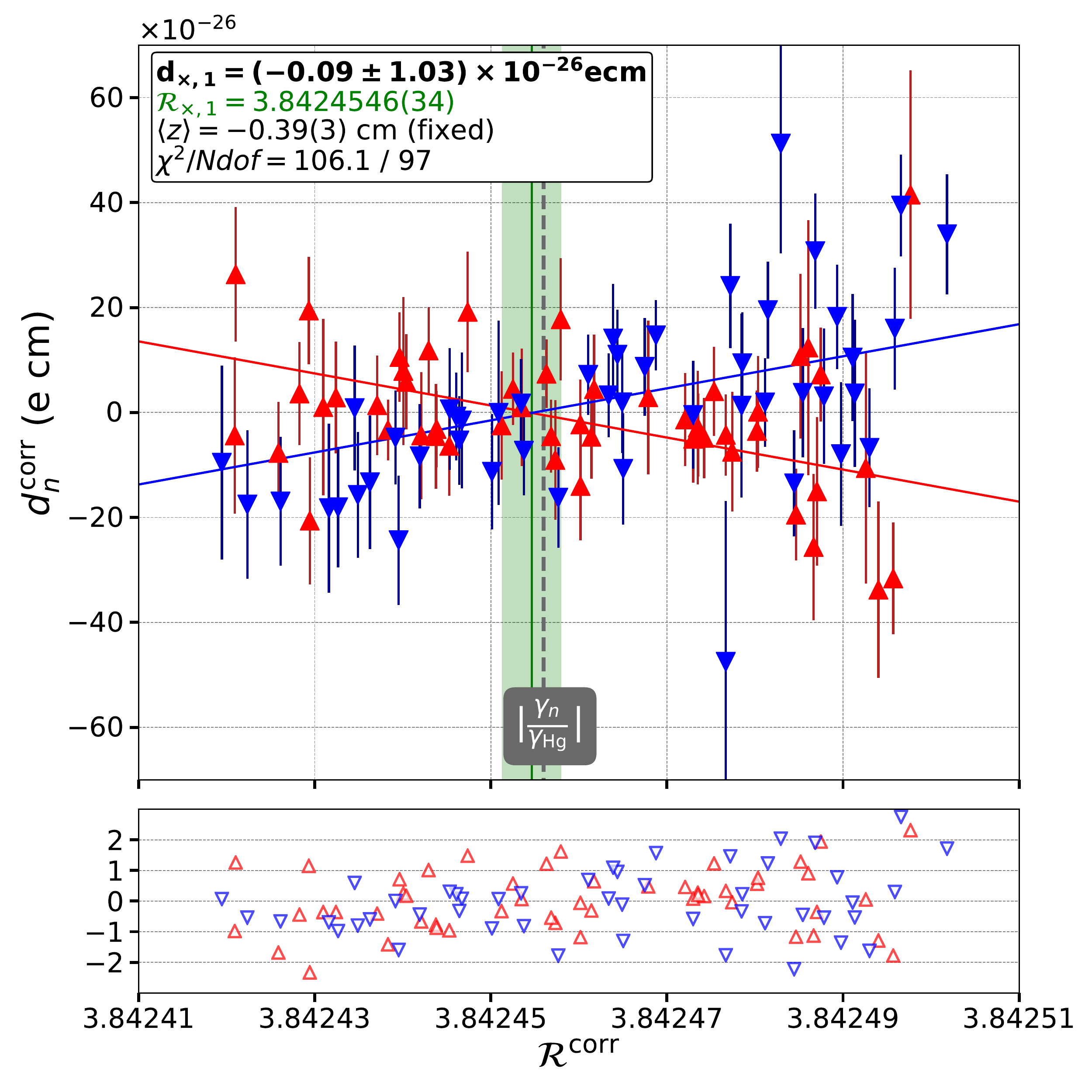}
\caption{Crossing point analysis: the corrected electric dipole moment $\dn^{\rm corr}$ is plotted versus $\R^{\rm corr}$ (see text for the exact definition of $\dn^{\rm corr}$ and $\R^{\rm corr}$). 
Upwards-pointing (red) and downwards pointing (blue) triangles represent sequences in which $B_0$ was pointing upwards and downwards, respectively. 
The fitted value of $\R_\times$ is represented by the green vertical band (1-$\sigma$), and the vertical dashed line represents the ratio of gyromagnetic ratios calculated from the literature values of $\gamma_\n$ \cite{Greene1979} and $\gamma_\Hg$ \cite{Cagnac1961}.  
The lower panel shows the normalized fit residuals. 
}
\label{fig:CrossingPoint}
\end{figure}

The dominant systematic effects arise from a shift linear in $E$ due to the combination of the relativistic motional field $\vec{B}_{\rm m} = \vec{E}\times\vec{v}/c^2$\,\cite{Pendlebury2004} and the magnetic field gradient: 
\begin{equation}
\delta_{\rm EDM}^{\rm false} = - \frac{2 E}{\hbar |\gamma_\n| B_0} (\dn^{\rm net} + d^{\rm false}), 
\end{equation}
where $\dn^{\rm net}$ is the effect of a possible net motion of the UCNs (discussed later) and $d^{\rm false}$ is due to the the random motion of the UCNs and \magHg{} atoms in a non-uniform magnetic field. 
The latter is largely dominated by the mercury and is written as~\cite{Pignol2012PRA, AbelPRA2019}: 
\begin{equation}
\label{eq:falseEDM}
d^{\rm false} = \frac{\hbar}{8c^2}\left|\gamma_\n \gamma_\Hg \right| R^2 \left(G_{\rm grav}+\hat{G}\right),
\end{equation} 
where $\hat{G}$ is the higher-order gradient term, which does not produce a gravitational shift. 
We used magnetic-field maps, measured offline, to extract a value of $\hat{G}$ for each sequence and calculate a corrected EDM value $\dn^{\rm corr} = \dn^{\rm meas} - \hbar\left|\gamma_\n \gamma_\Hg\right| R^2 \hat{G}/(8c^2)$. 
The main contribution in Eq.\,\eqref{eq:falseEDM} depending on $G_{\rm grav}$ is then dealt with by the crossing-point analysis, shown in Fig.\,\ref{fig:CrossingPoint}: 
$\dn^{\rm corr}$ is plotted as a function of $\R^{\rm corr} = \R/(1+\delta_{\rm T} + \delta_{\rm Earth})$, and we fit two lines with opposite slopes corresponding to the sequences with $B_0$ up and $B_0$ down. 
At the crossing point we have $G_{\rm grav} = 0$, and the main systematic effect is canceled. 
In the fit the free parameters are the coordinates of the crossing point $\R_\times$ and $d_\times$; the slope was fixed to the theoretical value $\partial d^{\rm false}/\partial \R =  \hbar\gamma_\Hg^2 R^2 B_0 / (8\langle z \rangle c^2) $. 
Because of the uncertainty on $\langle z \rangle = -\SI{0.39(3)}{cm}$, the slope has an error that propagates to become an additional error of $7 \times 10^{-28} \ecm$ on $d_\times$. 
As a check we also considered the slope as a free parameter in the fit and found $\langle z \rangle = \SI{-0.35 (6)}{cm}$, in agreement with the values found in Ref.\,\cite{AbelPRA2019}. 

In order to have $G_{\rm grav} = 0$ at the crossing point we had to correct $\R_i$ for all shifts other than the gravitational shift: 
namely the shift due to Earth's rotation $\delta_{\rm Earth}$, and the shift due to transverse fields $\delta_{\rm T} = \langle B_{\rm T}^2\rangle/(2B_0^2)$\,\cite{AbelPRA2019}. 
The transverse shift  for each sequence was calculated from the offline magnetic-field maps. 
The vertical corrections, related to $\hat{G}$, shifted the crossing point by $\unit[\pow{(69 \pm 10)}{-28}]{\ecm}$. 
The horizontal corrections, related to $\langle B_{\rm T}^2\rangle$, shifted the crossing point by $\unit[\pow{(0 \pm 5)}{-28}]{\ecm}$.

The corrections for the effect of the magnetic non-uniformities $\hat{G}$ and $\langle B_{\rm T}^2\rangle$ are based on the mapping of the apparatus without precession chamber, hence possibly missing the contribution of magnetic impurities in the precession chamber. 
All inner parts were scanned for magnetic dipoles before and after the data taking in the Berlin Magnetically Shielded Room-2 at the Physikalisch Technische Bundesanstalt in Berlin. 
%{\bf 
% why the bold font here??
Initially we verified, that all parts showed no signals above the SQUID system's detection threshold of $\SI{20}{nAm^2}$;
%} and 
the second scan revealed a dozen dipoles with a maximum strength of $\SI{100}{nA m^2}$. 
The corresponding systematic error was evaluated to be $\unit[4 \times 10^{-28}]{\ecm}$. 

In addition to the false EDM due to the random motional field $d^{\rm false}$, 
a net ordered motion of the UCN could generate a systematic effect $\dn^{\rm net} =\eta\epsilon\cdot\unit[\pow{6.7}{-23}]{\ecm/(m/s)}$, 
where $\eta$ is the mean net velocity of the ordered motion orthogonal to $E$ and $B$, and $\epsilon$ is the misalignment angle between the electric and magnetic fields. 
Three possible sources of ordered motion were identified in the past\,\cite{Pendlebury2015PRD}: a vertical motion due to micro heating, and initial transverse and rotational motions that are destroyed by collisions on the wall surfaces. Using the same trap geometry as in Ref.\,\cite{Pendlebury2015PRD} and a softer initial UCN spectrum\,\cite{Bison2019b}, we use the same value for $\epsilon$ and $\eta$. The error from heating was estimated to be $\unit[\pow{1}{-30}]{\ecm}$, while the error from rotational motion dominates: $\unit[\pow{2}{-28}]{\ecm}$.

The motional field also induces a shift quadratic in $E$ of  
$\delta_{\rm quad} = \gamma_\Hg^2 R^2 E^2 / (4 c^4)$ \cite{Pignol2015}, where we consider only the (dominant) shift on the mercury frequency. 
We were able to exclude any possible polarity dependence of the $E$-field magnitude to a level of $10^{-4}$ and therefore state
a conservative error of $\unit[\pow{0.1}{-28}]{\ecm}$ for this effect. 

Next, imperfect compensation of the $\delta_{\rm grav}$ term by the CsMs can lead to a direct systematic effect in case of a correlation between the $E$-field polarity and the magnetic-field uniformity. 
We evaluated the possible effect by de-activating the gradient drift compensation in both analysis and found mean difference of $\unit[\pow{7.5}{-28}]{\ecm}$; we quote the full shift as a systematic error. 
Leakage currents could be one source of such a correlation.

The term $\delta_{\rm light}$ corresponds to a mercury frequency shift proportional to the power of the UV probe light \cite{Cohen-Tannoudji1962}. 
We estimate that the largest shift of this type is at the level of $0.01$~parts per million in our experiment.
This can constitute a systematic effect if the power of the probe light is correlated with the polarity of the electric field, which we cannot exclude below the level of $0.14 \%$. 
This results in a systematic error of $\unit[\pow{0.4}{-28}]{\ecm}$ for mercury light shifts.

Ultracold neutrons co-precessing with polarized \magHg atoms are exposed to a pseudo-magnetic field $\vec{B}^\star = -4 \pi \hbar n_\Hg b_{\rm inc} \vec{P} \sqrt{1/3}/(m \gamma_\n)$\,\cite{Abragam} due to a spin-dependent nuclear interaction quantified by the incoherent scattering length $b_{\rm inc}(\magHg)=\pm\SI{15.5}{fm}$\,\cite{NistDataBase}. 
The mercury polarization $\vec{P}$ could have a residual static component $P_\parallel= |P|\sin\zeta$ in case of an imperfect $\pi/2$ pulse; this would generate a systematic effect if $P_\parallel$ correlates with the electric-field polarity. 
We deduced $\zeta$ from the photomultiplier signal of the probe beam during the $\pi/2$ flip. 
The product  $n_{\rm Hg}|P|$ was estimated by comparing the ratio of precession amplitude to total light absorption in the \magHg-lamp read-out  and matching this to a laser measurement to calibrate for a pure $\lambda = \SI{254.7}{nm}$ light source. 
The systematic error induced by the term $\delta_{\rm inc}$ was estimated to be $\unit[7 \times 10^{-28}]{\ecm}$.

Table \ref{tab:Systematic effects} lists the above-mentioned systematic effects. 
%We have also considered additional effects $\delta_{\rm other} =\delta_{\rm pulse}+ \delta_{\rm AC}$. 
% I commented out the above statement because \delta_other contains not only pulse and AC, but also Johnson, movement... And delta_pulse isn't a systematic so shouldn't be here.
Additionally, the mercury pulse causes a small tilt of the neutron spin prior to the Ramsey procedure, and is responsible for the shift $\delta_{\rm pulse}$. 
This shift is not correlated with the electric field; it behaves as an additional random error, and was accounted for in the statistical analysis. Further effects $\delta_{\rm other}$ that were also studied and found to be negligible (smaller than $\unit[10^{-29}]{\ecm}$) include: the effects of AC fields $\delta_{\rm AC}$ induced by ripple of the high voltage supply; noise of the current supplies, or Johnson-Nyquist noise generated by the electrodes; the movement of the electrodes correlated with electric field; and a correlation of the orientation of the magnetic field with the electric field in combination with the rotation  of the Earth. 

\begin{table}
\caption{Summary of systematic effects in $10^{-28}$\,{\ecm}. The first three effects are treated within the crossing-point fit and are included in $d_\times$. The additional effects below the line are considered separately.}
\label{tab:Systematic effects}

\begin{tabular}{lrr}
\hline
Effect & shift & error \\ \hline 
Error on $\langle z \rangle$ & - & 7 \\
Higher order gradients $\hat{G}$ & 69 & 10 \\
Transverse field correction $\langle B_{\rm T}^2\rangle$ & 0 & 5 \\
\hline
Hg EDM\cite{Graner2016PRL} & -0.1 & 0.1   \\
Local dipole fields  & - & 4 \\
$v \times E$ UCN net motion & - & 2 \\
Quadratic $v \times E$ & - & 0.1 \\
Uncompensated G drift & - & 7.5 \\
Mercury light shift & -  & 0.4 \\
Inc.\ scattering \magHg  & - & $7$\\
\hline \hline
TOTAL & 69 & 18 \\ \hline
\end{tabular}
\end{table}

During data-taking a copy of the files with the neutron detector data was modified by moving a predefined randomly distributed number of neutrons from one UCN detector to the other (see Fig.~\ref{fig:ramsey_example}). 
%The original files were encrypted (?) and stored on an access restricted server. 
This injection of an artificial EDM signal into the data was applied twice, and two datasets with different artificial EDMs were distributed to two distinct analyses groups\,\cite{KrempelEtAl}. 
This double-blind procedure enforced the independence of the two analyses, in particular for the data selection criteria. 
%Both groups developed their own analysis algorithms using different software and strategies; no code was exchanged.
%Each group independently evaluated criteria to cut or include data within the main data taking runs.
%Once both groups considered their analysis mature and ready for relative un-blinding the results were compared first before removing the second artificial EDM injection, and again after. 
Once the two analyses had been completed using only double-blind datasets, it was confirmed that they gave consistent results when run on an identical blind dataset. 
%. The two results have been compared by removing the differing part of the artifical offset. 
Finally both groups performed their analysis on the original never-blinded dataset. 
The results of the crossing-point fit are
$d_{\times,1}\!=\!\unit[\pow{(-0.09\pm1.03)}{-26}]{\ecm}$, $\R_{\times,1} = 3.8424546(34)$ with ${\rm \chi^2/dof}\!=\!106/97$  and  
$d_{\times,2}\!=\!\unit[\pow{(0.15\pm1.07)}{-26}]{\ecm}$, $\R_{\times,2} = 3.8424538(35)$ with ${\rm \chi^2/dof}\!=\!105/97 $.

%Our final result is the value with a better ${\rm \chi^2/dof}$ value. 
%We consider the second analysis as crosscheck which perfectly confirms the result. 
The small difference between the two results can be explained by the different selection criteria and we take as a final value the midpoint of the two. 
After adding the extra systematic effects quoted in the second part of Table \ref{tab:Systematic effects}, the final result, separating the statistical and systematical errors, is:
\begin{equation}
		\dn = \unit[\pow{(0.0\pm1.1_{\rm stat}\pm0.2_{\rm sys})}{-26}]{\ecm}.
\label{eq:finalResult}
\end{equation}

The result may be interpreted as an upper limit of $|\dn|< \unit[\pow{1.8}{-26}]{\ecm}$~(90\%~C.L.). 
This has been achieved through an unprecedented understanding and control of systematic effects in the experiment.  In particular, those related to magnetic-field nonuniformity were assessed with dedicated measurements that resulted in a significant correction, equivalent to 60\% of the statistical uncertainty, that arose from higher-order magnetic-field gradients. Overall the systematic error has been reduced by a factor of five compared to the previous best result \,\cite{Pendlebury2015PRD}. 

\section*{Acknowledgments}
%The dataset was taken in 2014 - 2017 at PSI Villigen. 
We are profoundly grateful for the fundamental contributions to the field in general and to this project in particular of J.M.~Pendlebury, the intellectual giant on whose shoulders we stand, and to K.F.~Smith and others also involved with the original development of the nEDM spectrometer with Hg co-magnetometer.  We acknowledge the excellent support provided by the PSI
technical groups and by various services of the collaborating
universities and research laboratories. 
In particular we acknowledge with gratitude the long term outstanding technical support by F.~Burri and M.~Meier. 
We thank the UCN source operation group BSQ for their support. We acknowledge financial support from the Swiss National
Science Foundation through projects 
No.\,117696,
No.\,137664,  
No.\,144473,
No.\,157079,
No.\,172626, 
No.\,126562,
No.\,169596~(all PSI), 
No.\,181996~(Bern),
No.\,162574~(ETH), No.\,172639~(ETH), and No.\,140421~(Fribourg). 
University of Bern acknowledges the support via the European Research Council under the ERC Grant Agreement No.\,715031-BEAM-EDM\@.  
Contributions of the Sussex group have been made possible via STFC grants ST/M003426/1, ST/N504452/1, and ST/N000307/1. LPC Caen and
LPSC Grenoble acknowledge the support of the French
Agence Nationale de la Recherche (ANR) under Reference
No.\,ANR-09-BLAN-0046 and the ERC Project No.\,716651-NEDM\@. The Polish collaborators acknowledge support
from the National Science Center, Poland, under Grants
No.\,2015/18/M/ST2/00056, 2016/23/D/ST2/00715 and 2018/30/M/ST2/00319. P.M.\ acknowledges Grant No.\,SERI-FCS 2015.0594. This work was also partly supported by
the Fund for Scientific Research Flanders (FWO), and Project
GOA/2010/10 of the KU Leuven.
We acknowledge the support
from the DFG (DE) specifically projects BI 1424/2-1 and /3-1.
In addition we are grateful
for access granted to the computing grid PL-Grid infrastructure.

%\bibliographystyle{NumEtAlNoTitle}
%\bibliography{nEDM-references,UCN-references,BSM-references,SM-references}

\input{nEDMLetterReferences}
\end{document}

%% file: packages.tex
\usepackage[T1]{fontenc}
\usepackage{aas_macros}
\usepackage{lmodern}
\usepackage[official,right]{eurosym}
\usepackage{textcomp}   % provides proper micro- and degree-symbol
\usepackage{gensymb}
\usepackage[english]{babel}
\usepackage[pdftex]{graphicx,epsfig}
\usepackage{rotating}
\usepackage{eso-pic}
\usepackage{amsmath,amsthm}
\usepackage{dsfont}\let\mathbb\mathds
\usepackage[section]{placeins} %allows to have unlimited floats
\usepackage{needspace}
\usepackage{setspace}
\usepackage[suffix='']{epstopdf}
\usepackage{tabularx}
\usepackage{multirow}
\usepackage{MnSymbol}
\usepackage[tight,nice]{units}
\usepackage{xspace}
\usepackage{pbox}
\usepackage{siunitx}%added by Martin

%% file: CommandsAndShortcuts.tex
\newlength{\bildtitel}
\newcommand\REVIEW[1]{\message{LaTeX Warning: \noexpand untreated nEDM-REVIEW command in \jobname .tex: l\the\inputlineno}}% for publication
\newcommand{\diff}[1]{\operatorname{d}\ifthenelse{\equal{#1}{}}{\,}{\!#1}}

\newcommand{\pow}[2]{\ensuremath{#1\!\times\!10^{#2}}}

\newcommand{\ecm}{\ensuremath{\si{\elementarycharge}\!\cdot\!\cm}}
\newcommand{\dn}{\ensuremath{d_\text{n}}}

\newcommand{\nT}{\mbox{nT}}

\newcommand{\cm}{\ensuremath{\mathrm{cm}}}

\newcommand{\magHg}{\ensuremath{{}^{199}\text{Hg}}}